# THERMAL CONDUCTIVITY OF HEMP CONCRETES: VARIATION WITH FORMULATION, DENSITY AND WATER CONTENT.

## Florence COLLET [a] [*], Sylvie PRETOT [a]

[a] Université Européenne de Bretagne - Laboratoire de Génie Civil et Génie Mécanique – Equipe Matériaux Thermo Rhéologie

Postal address : IUT Génie Civil – 3, rue du Clos Courtel – BP 90422 – 35704 Rennes – France

florence.collet@univ-rennes1.fr

sylvie.pretot@univ-rennes1.fr

[*] Corresponding author: Tel: 33.2.23.23.40.56, Fax: 33.2.23.23.40.51.

## Abstract

This study investigates the effect of formulation, density and water content on the thermal conductivity of hemp concretes. The investigations are based on experimental measurements and on self-consistent scheme modelling. The thermal conductivity of studied materials ranges from 90 to 160 mW/(m.K) at (23°C; 50%HR). The impact of density on thermal conductivity is much more important than the impact of moisture content. It is shown that the thermal conductivity increases by about 54 % when the density increases by 2/3 while it increases by less than 15 % to 20 % from dry state to 90%RH.

## 1    Introduction

In a context of sustainable development, green buildings aim at reducing the environmental impacts of buildings while also ensuring high indoor environmental quality (comfortable and healthy). The main impacts of buildings on the environment are due to consumption of resources (energy, raw material,

water…) and to emissions (greenhouse gas, pollution, wastes...). Thus, green buildings should be energy efficient while showing a light foot-print on the environment over the entire life-cycle.

The energy efficiency of buildings depends on the hygrothermal behaviour of the building envelope and on the performance of systems. This behaviour is related to hygric and thermal properties of constitutive materials. Among these properties, the thermal conductivity is dependent on several parameters such as density, water content and temperature. The bibliography provides studies on the effect of density, of moisture content and of temperature on the thermal conductivity of building materials. For example, Uysal and al. studied the effect of cement dosage and of prumice aggregate ratio on density and thermal conductivity of concrete [Uysal and al., 2004]. In all studied cases, the thermal conductivity increases with density. The prumice aggregate allows reducing the thermal conductivity by 46% while reducing the density by 40%. Del Coz Díaz and al. measured the thermal conductivity of lightweight concrete produced from expanded clay. The thermal conductivity increases from 176 to 256 mW/(m.K) when the density increases from 973 to 1362 kg/m$^3$ at 23°C, 50% RH. Bederina and al. have shown that the addition of wood shaving in dune sand concrete allows reducing the thermal conductivity from 1.20 to 0.55 W/(m.K) while reducing the density from 2100 to 1400 kg/m$^3$ [Bederina and al., 2006]. In addition, even if literature mainly provides dry state values, the thermal conductivity of hygroscopic materials increases with moisture content. For example, from dry state to saturated state, (*i*) the thermal conductivity of autoclaved aerated concrete increases up to six times [Jerman and al., 2013], (*ii*) the thermal conductivity of wood-concrete composite increases up to 2.76 times [Taoukil and al., 2013]. For lightweight concrete, the thermal conductivity increases by 2 from dry state to 100% RH [del Coz Díaz and al., 2013]. These authors give mathematical expression which defines the relationship between density, moisture content and thermal conductivity. Finally, Jerman and al. show that the dependence of thermal conductivity on temperature is less important than its dependence on moisture content for temperatures between 2 and 30°C [Jerman and al., 2013].

Bio-based building materials are derived from renewable biological resources. So, they are an answer to the problem of resource depletion. They are drawn from various raw materials such as wood [Aigbomian and al., 2013], coconut [Alavez-Ramirez and al., 2012], diss [Sellami and al., 2013], sunflower [Binici and al., 2013], hemp... Hemp straw provides two products used in building materials. Fibbers are used to produce insulating panels [Arnaud, 2000] [Collet and al., 2011] and shiv are used as bio-aggregate in hemp composites (concrete, render…). Hemp concrete is used for several applications: wall, floor or roof. The main difference between applications is the hemp to binder ratio in the mix. Furthermore, hemp concrete can be produced by spraying, moulding or precasting.

Hemp concrete is environmentally friendly as it is made of a renewable raw material (hemp) and as it allows carbon sequestration [Boutin and al., 2005][Ip and Miller, 2012][Pretot and al., 2014]. Hemp concrete is highly porous, with open and interconnected porosity. It is a lightweight material with densities between 200 and 600 kg/m$^3$, depending on application [Amziane and Arnaud, 2013]. Therefore, it shows quite low thermal conductivity, about 100 mW/(m.K) for a walling application [Arnaud, 2000] similar to other building materials with comparable density (cellular concrete for example). The thermal conductivity of hemp concrete depends on the amount of shiv which impacts density. For hemp percentage in weight between 20 and 40%, the density decreases from 611 to 369 kg/m$^3$ and the thermal conductivity decreases from 140.8 to 94.7 W/(m.K) [Benfratello and al., 2013]. The open and interconnected porosity allows moisture transfer and storage. The hygric characteristics of the material are moisture permeability and sorption curves that are representative of steady state. Previous studies have shown that hemp concrete is strongly hygroscopic, with high moisture transfer and storage capacities (water vapour permeability about 3.2 E-11 kg/(m.s.Pa) and sigmoid sorption curves with high hysteresis loop between adsorption and desorption curves) [Collet and al., 2008] [Collet and al., 2013]. These properties allow hemp concrete to moderate ambient relative humidity variations. This ability is quantified through the Moisture Buffer Value of the material. Studies performed following the Nordtest Project classify hemp concrete from good to excellent hygric regulator (MBV around 2 g/(m².%RH)) [Rode, 2005][Collet and Pretot, 2012][Collet and al., 2013]. For comparison, the MBV of cellular concrete is about 1 g/(m².%RH) [Rode, 2005]. Studies performed at wall, or building scale have highlighted sorption / desorption or condensation / evaporation phenomena in hemp concrete wall [Pretot and Collet, 2012][Shea and al., 2012]. These phenomena involve binding or latent energy that impact energy balance of the wall. Thus, hemp concrete shows high hygrothermal performances which allow energy saving and high indoor comfort [Tran Le and al., 2010] [Evrard and de Herde, 2010]. For example, Tran Le and al. performed a numerical study to compare hemp concrete behaviour to that of cellular concrete. They found that hemp concrete induces a reduction ranging from 15 to 45 % in energy consumption, depending on ventilation strategy.

Actually, the energy needs (and indoor comfort) can be simulated thanks to dynamic thermal modelling which requires, among input data, the hydric and thermal properties of building materials. Such numerical models should take into account the variation of hydric and thermal properties with hygrothermal state like the variation of hydric properties with temperature or the variation of thermal properties with humidity.

This study deals with the thermal conductivity of five hemp concretes that differ from formulation. Firstly, the thermal conductivity of hemp concretes is studied versus formulation. The impact of raw materials and hemp to binder ratio is highlighted. Then, the increase of thermal conductivity with dry density and water content is considered. The investigations are based on experimental measurements and on self consistent scheme modelling.

## 2    Materials

### 2.1    Formulations and production methods

Hemp concrete is a bio-aggregate-based building material made of hemp shiv and binder. This study investigates five hemp concretes that differ from formulation and production method. This work was supported by two industrial partners. The studied materials are representative of the materials usually produced by the partners.

The spraying method consists in mixing hemp shiv and lime-based binder to constitute a dry mix that is blown along a pipe by means of a flow of compressed air. Water is added just before the hose outlet, the quantity of which may be controlled by the operator via one valve. This allows reducing the quantity of water compared to conventional mixing which requires more water due to the absorption capacity of hemp shiv. In this study, hemp concrete is sprayed into moulds (figure 1). Three hemp to binder ratios are considered regarding practices: wall, floor and roof (table 1). Moreover, the spraying method induces small density variation due to the angle and the distance of spraying [Elfordy and al., 2008], so three densities of sprayed hemp concrete are considered (light, medium and heavy) for the wall mix.

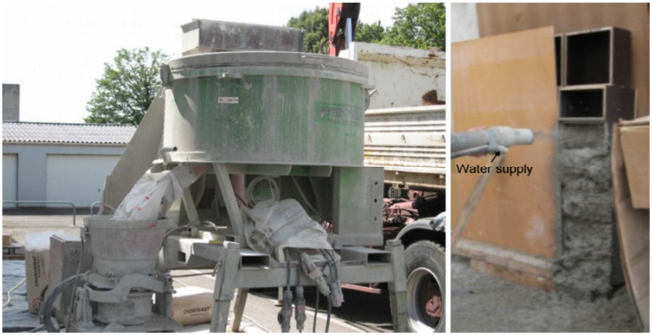

**Figure 1. Spraying method (left : mixer, right : spraying hose)**

The moulding method consists in mixing hemp shiv and lime-based binder in a mixer. Water content is adjusted to obtain a consistent workability of fresh hemp concrete. Moulds are filled with the mixture and the hemp concrete is slightly compacted.

The precasting method results from an industrial process. Firstly, slaked lime is produced from CaO and water. Hydraulic lime and hemp shiv are then added and the mixture is poured into moulds. Blocks are finally formed by compaction under vibrations (figure 2).

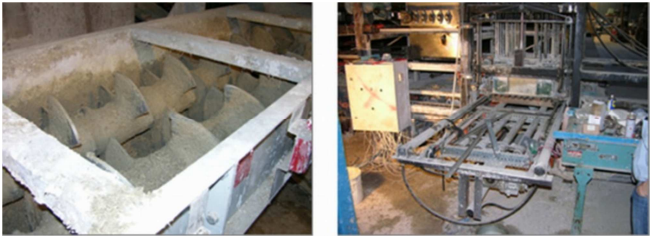

**Figure 2. Precasting method (left : mixer, right : block press)**

For this study, for all formulations and production methods, hemp concrete is manufactured in moulds of 30 cm×30 cm×16 cm (figure 1 and 2). This size is representative of a wall thickness or of industrial precast blocks. Once produced, blocks are stabilized at 23°C, 50% RH. Then, the blocks are cut to obtain 15×10×5 cm$^3$ specimen. For SHC roof and for PHC, it was not possible to cut the blocks because they were exfoliating. For these materials the measurement are performed on blocks at (23°C; 50%RH) only.

Table 1 summarises the mix proportioning and manufacturing method of studied hemp concretes:

Sprayed Hemp Concrete (SHC), Moulded Hemp Concrete (MHC) and Precast Hemp Concrete (PHC).

**Table 1: Mixes proportioning and manufacturing method of studied hemp concretes**

| Material | Method of Production | Binder | Hemp shiv | Hemp/Binder mass ratio |
|---|---|---|---|---|
| SHC Wall : Sprayed Hemp Concrete - wall | Spraying | Commercial lime-based binder * | Commercial defibered hemp shiv | 0.5 |
| SHC Floor : Sprayed Hemp Concrete - floor | Spraying | Commercial lime-based binder* | Commercial defibered hemp shiv | 0.4 |
| SHC Roof : Sprayed Hemp Concrete - roof | Spraying | Commercial lime-based binder* | Commercial defibered hemp shiv | 1 |
| MHC : Moulded Hemp Concrete (wall) | Moulding | Commercial lime-based binder* | Commercial Fibered Hemp Shiv | 0.5 |
| PHC : Precast Hemp Concrete (wall) | Precasting | CaO : 72% Hydraulic lime : 28% | Defibered hemp shiv | 0.65 |

*lime-based binder made of 75% of hydrated lime (98% CaO), 15 % of hydraulic binder and 10 % of pozzolanic binder*

## 2.2    Density and porosity

Density is calculated from mass and dimensions of specimens. Porosity is measured by pycnometry. Table 2 gives the apparent density at (23°C, 50%RH) and the total porosity of studied materials.

For sprayed hemp concretes, the apparent density ranges from 260 kg/m$^3$ for roof to 460 kg/m$^3$ for floor. For wall mix, it ranges from 390 kg/m$^3$ for light wall to 460 kg/m$^3$ for heavy wall. The moulded hemp concrete shows an apparent density of 381 kg/m$^3$ while the precast one shows an apparent density of 457 kg/m$^3$. These values are consistent with the values commonly found in literature [Amziane and Arnaud, 2013].

The total porosity of studied hemp concretes ranges from 72% for precast hemp concrete to 85% for sprayed hemp concrete roof. This high porosity includes a wide range of pores from micrometric pores in binder matrix and hemp shiv to millimetric pores due to the arrangement between the hemp shiv and to the hemp–binder adhesion [Collet and al., 2008].

**Table 2: Density and porosity of studied hemp concretes**

| Material | Apparent density at (23°C; 50%RH) [kg.m⁻³] | Total porosity n [%] |
|---|---|---|
| SHC lw : Sprayed Hemp Concrete – light wall | 390-401 | 79.0-79.5 |
| SHC mw : Sprayed Hemp Concrete – medium wall | 420-426 | 77.8-78.0 |
| SHC hw : Sprayed Hemp Concrete – heavy wall | 446-463 | 75.7-76.6 |
| SHC Floor : Sprayed Hemp Concrete - floor | 460 | 78.7 |
| SHC Roof : Sprayed Hemp Concrete - roof | 258 | 84.9 |
| MHC : Moulded Hemp Concrete (wall) | 381 | 84.3 |
| PHC : Precast Hemp Concrete (wall) | 457 | 72 |

# 3 Methods

## 3.1 Experimental investigation of thermal conductivity

### 3.1.1 Hot wire method

The thermal conductivity $\lambda$ ($W.m^{-1}.K^{-1}$) was measured using the commercial CT-meter device. In this study, this device was equipped with a five-centimeter-long hot wire (figure 3). The measurement is based on the analysis of the temperature rise versus heating time. For cylindrical geometry, Blackwell [Blackwell, 1953] and Carslaw and Jaeger [Carslaw and Jaeger, 1959] solve the equation of heat conduction for a two media system including (*i*) the probe, assumed as an ideal infinitely thin and long line heating source, and (*ii*) the studied material, that constitutes an infinite surrounding and is supposed to be homogeneous and isotropic. For a sufficiently long time, there is a proportional relationship between temperature rise $\Delta T$ and logarithmic heating time ($\ln(t)$) (figure 3)(1):

$$\Delta T = \frac{q}{4.\pi.\lambda}(\ln(t) + K) \qquad (1)$$

Where $q$ is the heat flow per meter (W.m$^{-1}$) and K is a constant including the thermal diffusivity of the material.

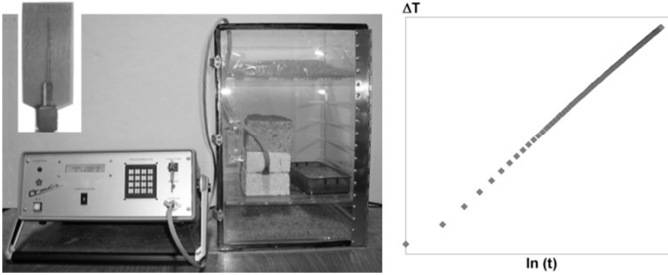

**Figure 3. Measurement of thermal conductivity at dry state – left : CT Meter, Hot wire and dry chamber, right : experimental thermogram**

The heat flow and heating time are chosen to reach high enough temperature rise (>10°C) and high correlation coefficient (R²) between experimental data and equation (1). In this study, the heat flow ranges from 2.9 to 4.6 W/m and the heating time is 120 seconds. According to the manufacturer, the hot wire is well adapted for the measurement of thermal conductivities ranging from 0.02 to 5 W.m$^{-1}$.K$^{-1}$ and the expected accuracy is 5%.

The main advantage of this method, compared to steady-state methods like hot plate, is that it is a transient method that does not induce (or that does limit) water migration during test [Hladik, 1990]. This allows the study of the effect of moisture content on thermal conductivity. The main disadvantage is that it is a localized measurement. Thus, measurements should be taken several times to ensure the representativeness of the thermal conductivity value. In this study, the given thermal conductivity is the average of five values with a variation coefficient lower than 5%. In addition, the correlation coefficient R² between experimental data and equation (1) is higher than 0.999 for all the measurements.

An infrared photography (figure 4), taken immediately after the measurement, allows viewing the cross section of the probed volume. The thermal wave doesn't reach the edges of the sample, so the hypothesis of infinite surrounding assumed in the model of the hot wire is valid. The radius of the probed volume is about 2.5 cm (figure 4), thus the probe volume is higher than the representative volume of hemp concrete (representative length about 5cm) [Collet and al., 2013].

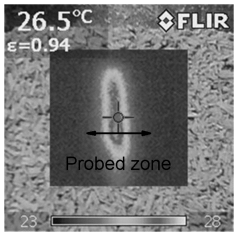

**Figure 4. Infrared photography of the probed area, immediately after measurement.**

### 3.1.2 Conditioning of specimens

In order to measure the effect of moisture content on thermal conductivity, the specimens are conditioned at 23°C to several ambient relative humidity (dry point, 33%RH, 50% RH, 81%RH and 90%RH) using dry chamber or climatic chamber Vötsch VC4060. Prior to the test, the specimens are discontinuously weighed until a constant mass is reached (difference lower than 0.01% for two consecutives weighing with 48 hours time step). The water content by mass w (g/g) is then calculated from wet mass and dry mass of the specimen.

During the test, specimens are kept at constant relative humidity in dry or climatic chamber.

## 3.2 Modelling thermal conductivity of hemp concrete

### 3.2.1 Modes of heat transfer in hemp concrete

In porous media, heat transfer can take place by three modes such as conduction, convection and radiation.

As soon as there is a temperature gradient, conduction takes place by means of molecular agitation within the material. The transfer occurs from higher temperature area to lower temperature area and acts to equalize temperature differences. Conduction takes place in all forms of matter such as solid, liquid and gas.

Heat transfer by convection results from convective currents of fluid that transfer heat from hotter place to cooler place. According to [Saatdjian, 1998], free convection takes place if the Rayleigh number $Ra$ is higher than $4\pi^2$. The Rayleigh number is the product of the Prandtl number $Pr$ and the Grashof number $Gr$. The Prandtl number is the ratio of momentum diffusivity (kinematic viscosity) to thermal diffusivity; it is a characteristic of the fluid itself. The Grashof number approximates the ratio of the buoyancy to viscous force acting on a fluid (2) :

$$Gr = \frac{g \times \beta \times \rho^2}{\mu^2} \Delta\theta \times L^3 \qquad (2)$$

With $(i)$ $g$ the gravitational acceleration [m/s$^2$], $(ii)$ $\beta = \frac{1}{T}$ the coefficient of thermal expansion [K$^{-1}$], $(iii)$ $\rho$ the density of air [k/.m$^3$], $(iv)$ $\mu$ the dynamic viscosity [kg/(m.s)], $(v)$ $L$ the maximum length of pores [m] and $(vi)$ $\Delta\theta$ the difference between the temperature of the surface and the temperature of the fluid [°C], estimated from the temperature rise during the measurement $\Delta\theta_{max}$ (about 10°C), the radius of the probed volume $r$ (2.5 cm) and the maximum length of pores $L$ (3mm) :

$$\Delta\theta = \frac{\Delta\theta_{max}}{r} \times L \qquad (3)$$

For a temperature of 300 K, the density of air $\rho$ is 1.177 kg/m$^3$, the dynamic viscosity $\mu$ is 1.85E-05 kg/(m.s) and the Prandtl number is 0.708 [Sacadura, 1978]. The Rayleigh number is thus equal to 3.04. This value is lower than $4\pi^2$ for the maximum length of pores, so the free convection doesn't occur in hemp concrete.

Lastly, all matter with a temperature greater than absolute zero emits thermal radiation that is a consequence of thermal agitation of its composing molecules. The radiative effect in hemp concrete during the measurements is estimated from the model of Siegel and Howel [Siegel and Howel, 1972]. This model gives the radiative heat flux assuming that the porous medium is a set of parallel screens

(solid phase) separated by a fluid phase (with absorption coefficient $\alpha_f$). Gliksman, according to [Kalboussi, 1990], gives an equivalent thermal conductivity to radiative transfers considering that the fluid phase is non absorbent ($\alpha_f$=0), that the difference of temperature between the two sides of the sample is quite low, and that the distance between the screens is equal to the mean diameter of pores (4).

$$\lambda_{rad} = \frac{4\sigma \times D_{mean} \times T_{av.}^3}{\frac{2}{\varepsilon} - 1} \tag{4}$$

Where (i) $\sigma$ is the Stefan–Boltzmann constant (5.67 E$^{-08}$ W/(m².K$^4$)] ; (ii) $D_{mean}$ is the mean diameter of pores estimated to 1.2 E-05 m from mercury porosimetry results and from millimetric porosity of hemp concrete [Collet, 2004][Collet et al., 2008] ; (iii) $T_{av.}$ is the average temperature between the fluid and the solid phase (about 301 K) ; (iv) $\varepsilon$ is the emissivity, estimated to 0.9.

The thermal conductivity equivalent to radiative transfer is thus about 0.06 mW.m$^{-1}$.K$^{-1}$. This value is much lower than the thermal conductivity of hemp concrete (higher than 100 mW.m$^{-1}$.K$^{-1}$), so the radiative transfer is negligible in hemp concrete.

Finally, the heat transfer in hemp concrete can be considered as only conductive. So, a conductive model can be used for modelling the thermal conductivity of hemp concrete.

### 3.2.2 Self-consistent scheme

The self consistent scheme allows the determination of macroscopic physical properties of heterogeneous materials. It was developed for the mechanical characterization [Hill, 1965] [Wu, 1966] and was then extended to electrostatics, magnetostatics, electric conduction and thermal conduction that are mathematically analogous [Landauer, 1952] [Kerner, 1956] [Hashin, 1968].

The heterogeneous medium, with spherical inclusions, is assumed to be an assembly of spherical composite spheres of various sizes (figure 5). For a two-phase medium, a sphere of radius $R_a$ (phase a) is embedded in a concentric spherical shell of external radius $R_b$ (phase b). This composite sphere is embedded in a homogeneous and isotropic equivalent medium (figure 6).

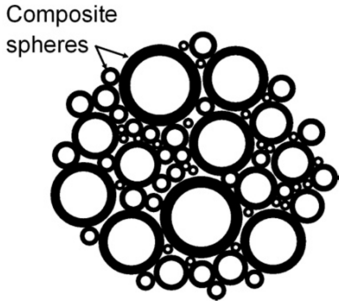

Composite spheres

**Figure 5. Heterogeneous medium assumed to be an assembly of spherical composite spheres of various sizes.**

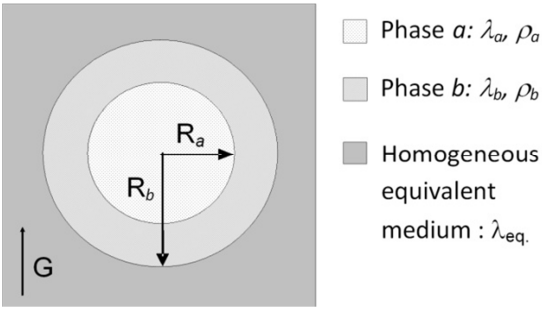

☐ Phase $a$: $\lambda_a$, $\rho_a$

☐ Phase $b$: $\lambda_b$, $\rho_b$

☐ Homogeneous equivalent medium : $\lambda_{eq.}$

**Figure 6. Two-phase composite sphere embedded in a homogeneous equivalent medium.**

The approximation of the effective conductivity is based on the assumption that the energy embedded in the heterogeneous medium is equivalent to that of the equivalent homogeneous medium under the same boundary conditions. The effective conductivity is thus linked to the volume ratio n of the inner to the outer radius according (5):

$$\lambda_{eq.} = \lambda_b \times \left( 1 + \frac{n}{\left( \frac{1-n}{3} + \frac{1}{\lambda_a/\lambda_b - 1} \right)} \right) ; n = \left( \frac{R_a}{R_b} \right)^3 \qquad (5)$$

Where $\lambda_{a\,(resp.b)}$ is the thermal conductivity of the phase a (resp. b).

For a three-phase medium, taking into account a supplementary shell of external radius $R_c$(phase c), this expression becomes [Boutin, 1996](6):

$$\lambda_{eq.} = \lambda_c \times \left( 1 + \frac{\varepsilon}{\frac{1-\varepsilon}{3} + \frac{3+\delta\left(\lambda_a/\lambda_b - 1\right)}{3\left(\lambda_a/\lambda_c - 1\right) - \delta\left(\lambda_a/\lambda_b - 1\right)\left(2\lambda_b/\lambda_c + 1\right)}} \right) ; \varepsilon = \left(\frac{R_b}{R_c}\right)^3 \text{ and } \delta = 1 - \left(\frac{R_a}{R_b}\right)^3 \tag{6}$$

In this study, the two phase model is used for dry material considering air as phase $a$ and solid matrix (including hemp shiv and binder) as phase $b$, $n$ is thus the porosity of the material. The three phase model is used for wet material considering air as phase $a$, water as phase $b$ and solid phase as phase $c$. $\varepsilon$ is the porosity of the material and $\delta$ is calculated from water content and porosity. The thermal conductivity of the solid phase is calculated by fitting the experimental data with the self consistent scheme using the least square method. This value is then used to compute the variation of thermal conductivity with density of dry material, and the variation of thermal conductivity with water content of wet material.

## 4    Results and discussion

### 4.1    Variation of thermal conductivity of Hemp Concretes versus formulation at (23°C; 50%RH).

Figure 7 gives the thermal conductivity of the five studied hemp concretes versus density at (23°C; 50%RH). For the wall formulations, the thermal conductivity ranges from 120 to 150 mW/(m.K). For sprayed hemp concrete walls (diamonds, triangles and squares on figure 7), the variation of thermal conductivity versus density fit a linear regression curve.

The SHC roof shows the lowest thermal conductivity, 93 mW/(m.K), mainly due to its lowest density. The SHC floor shows the highest thermal conductivity: 157 mW/(m.K). These values are not in line with the fitting curve of SHC wall. Actually, the hemp to binder ratio of SHC roof is higher than the hemp to binder ratio of SHC wall; and inversely the hemp to binder ratio of SHC floor is higher than the hemp to binder ratio of SHC wall.  So, as it was shown in [Benfratello and al., 2013], the thermal conductivity of hemp concrete depends on the amount of shiv and is not linearly proportional to the increase of shiv in mixture. The amount of shiv actually impacts the thermal conductivity of the solid phase of hemp concrete.

The thermal conductivity of moulded hemp concrete is higher than the value given by the fitting curve of SHC wall. As moulded hemp concrete has a higher porosity, its higher conductivity must be due to higher conductivity of its solid phase, related to higher real density. This may also be due to a higher conductivity of fibered hemp shiv, compared to defibered hemp shiv, as the fibers are less porous than the shiv and are thus more conductive.

The thermal conductivity of precast hemp concrete is lower than the value given by the fitting curve of sprayed hemp concrete wall. On one hand, the hemp to binder ratio of precast hemp concrete is higher than the hemp to binder ratio of sprayed hemp concrete, like previously this induces a lower thermal conductivity. On the other hand, the type of binder also impacts the thermal conductivity of materials. It is shown in [Gourlay and Arnaud, 2010] that the thermal conductivity of hydraulic lime binder is higher than the thermal conductivity of hydrated lime binder. Furthermore, Stefanidou and al. [Stefanidou and al., 2010] underline that adding white cement increases the thermal conductivity of mixtures and that adding pozzolanic materials reduces the thermal conductivity in comparison with pure lime.

Finally, the values measured on wet hemp concretes (at 23°C; 50%RH) are in agreement with the values found in literature. Actually, de Bruijn and Johansson [de Bruijn and Johansson, 2013] studied the thermal conductivity of two lime-hemp mixes at 15%RH and 65%RH. At 65%RH, they give thermal conductivity values of 116 and 100 mW/(m.K) when the densities are respectively 394.8 and 298.1 kg/m$^3$. Evrard [Evrard, 2008] gives a thermal conductivity ranging from 122 to 137 mW/(m.K) in moist state for density about 460 kg/m$^3$. According to [Elfordy and al., 2008], the thermal conductivity of sprayed hemp concrete is 179 mW/(m.K) for a density of 417 kg/m$^3$. This is higher than the values found in this study. This may be due to the moist state that is not well known in [Elfordy and al., 2008].

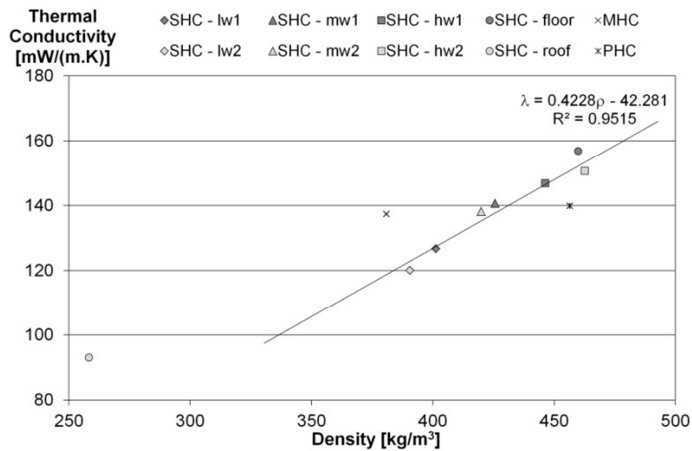

**Figure 7. Thermal conductivity of hemp concretes versus density at (23°C; 50%RH), for sprayed hemp concretes SHC (walls, floor, roof), moulded hemp concrete MHC and precast hemp concrete PHC - experimental data and linear fit of sprayed hemp concrete walls**

## 4.2  Variation of thermal conductivity of dry hemp concretes with density

Figure 8 gives the variation of thermal conductivity with the density of dry hemp concretes (sprayed hemp concretes: wall and floor and moulded hemp concrete). The experimental data of thermal conductivity of SHC wall ranges from 116 to 125 mW/(m.K) while the density ranges from 374 to 416 kg/m$^3$. The thermal conductivity of moulded hemp concrete is 11% higher: 129.6 mW/(m.K) for $\rho$ = 377 kg/m$^3$ . Lastly, the thermal conductivity of SHC floor is equal to 145 mW/(m.K) for a density of 450 kg/m$^3$. Compared to the values found in literature, on the one hand, these results are higher than the values obtained from the relationship established by Cerezo [Cerezo, 2005](eq. 7), that gives a thermal conductivity of 99.4 mW/(m.K) when the density is 400 kg/m$^3$.  On the other hand, these results are in the range of the values given by Nguyen [Nguyen, 2010] for similar density.  Cerezo studied hemp concrete manufactured by casting while Nguyen studied hemp concrete manufactured by compacting. It seems that the spraying method induces thermal conductivity similar to the values obtained from compaction.

$$\lambda = 0.0002 \times \rho + 0.0194 \qquad\qquad (7)$$

The thermal conductivity of the solid phase of hemp concretes is calculated following the method given in section 3.2.2. The thermal conductivity of the solid phase is equal to 61.2 mW/(m.K) for the SHC wall while it is equal to 95.3 mW/(m.K) for the MHC and to 79.1 mW/(m.K) for the SHC floor. As assumed previously, the solid phase of moulded hemp concrete is more conductive than the solid phase of sprayed hemp concrete wall, related to the use of fibered hemp shiv. The solid phase of the SHC floor is 30 % more conductive than the solid phase of the SHC wall, in line with lower hemp to binder ratio.

The self consistent scheme models the increase of thermal conductivity with density. For SHC wall, the theoretical result fits well the experimentation. This allows validating the numerical model, which is supposed to be valid for SHC floor and MHC to. When the density increases by 2/3, from 300 to 500 kg/m$^3$, the thermal conductivity increases from 97.1 to 149.2 mW/(m.K) for the SHC wall, from 107.6 to 165.9 mW/(m.K) for the MHC and from 103.4 to 159.3 mW/(m.K) for the SHC floor. So, for all these hemp concretes, the thermal conductivity increases in the same way with density (by about 54 % when the density increases by 2/3).

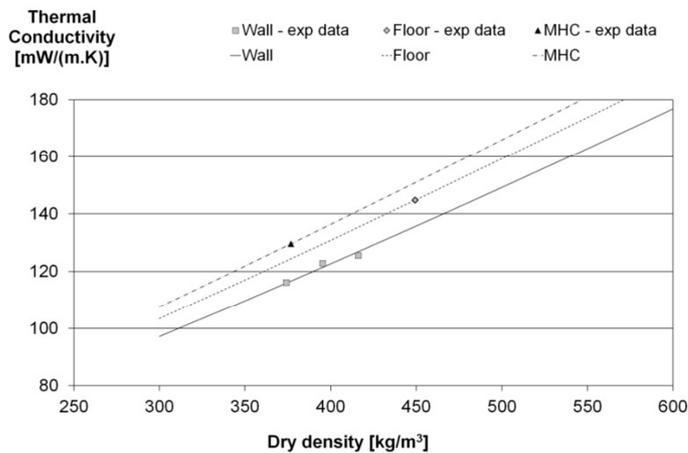

**Figure 8.  Thermal conductivity of hemp concretes versus density at dry state for sprayed hemp concretes (wall and floor) and moulded hemp concrete - experimental data (points) and two-phase self-consistent scheme (lines).**

**4.3    Variation of thermal conductivity of hemp concretes with water content**

Figures 9 and 10 give respectively the variation of thermal conductivity with water content for several density of SHC wall and for MHC and SHC floor. The points are experimental data and the lines model the variation of thermal conductivity with water content according the three phase self consistent scheme (equation 6). In all cases, there is a very high correlation between experimental data and self consistent scheme.

For the light SHC wall (dry density about 364 kg/m$^3$), the thermal conductivity increases by 25 % from 107 mW/(m.K) at dry state to 134.3 mW/(m.K) at 0.16 g/g in water content (reached at 90 %RH). For the medium and the heavy SHC wall, the thermal conductivity shows similar increase with water content.

The MHC and the SHC floor also show a similar increase of thermal conductivity with water content. From dry state to 0.05 g/g in water content (reached around 81%RH), the thermal conductivity rises respectively by 6.3% and 6.9%.

In the region of highest relative humidity, the thermal conductivity of MHC and SHC floor are less impacted than the thermal conductivity of SHC wall due to their sorption curves that show lower water contents [Collet and al., 2013].

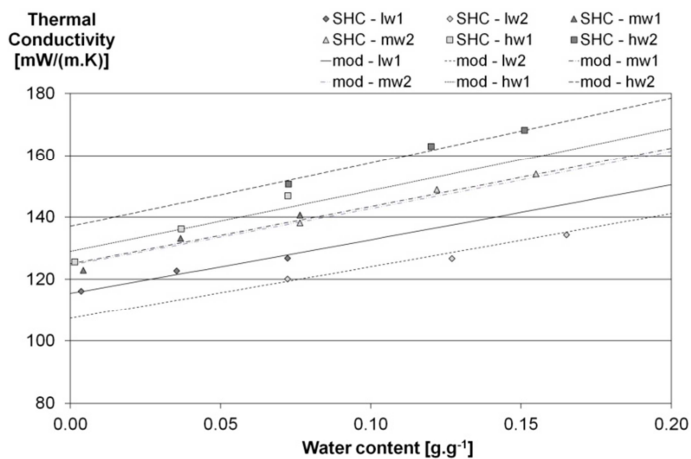

**Figure 9. Thermal conductivity of hemp concretes versus water content for sprayed hemp concrete walls (light, medium and heavy) - experimental data and three phase self consistent scheme.**

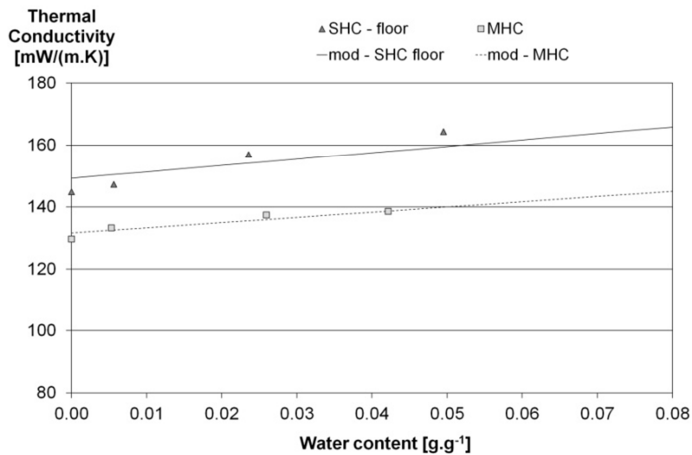

**Figure 10. Thermal conductivity of hemp concretes versus water content for sprayed hemp concrete floor and for moulded hemp concrete - experimental data and three phase self consistent scheme.**

## 4.4    Synthesis – Variation of thermal conductivity of hemp concretes versus dry density and water content

Figure 11 and table 3 summarise the variation of thermal conductivity of SHC wall with dry density and water content.

According to [Collet and al., 2013], the water content of SHC wall (with a dry density around 400 kg/m$^3$) is lower than 0.10 g/g for a wide range of relative humidity (up to 90%RH). From dry state to 0.10 g/g in water content, the thermal conductivity of SHC wall increases by 12.8% for low density (250 kg/m$^3$) and by 17% for high density (600 kg/m$^3$). For highest values of water content (0.20 g/g), the increase of thermal conductivity from dry state is 25.7% for low density and 34.8% for high density.

Moreover, from low density (250 kg/m$^3$) to high density (600 kg/m$^3$) the thermal conductivity increases by 109% at dry state and by 117% at 0.10 g/g in water content.

So, the impact of density on thermal conductivity is much more important than the impact of moisture content. Similar results are obtained with MHC and SHC floor. However, once the hemp concrete is produced, its density remains the same while its water content varies according surrounding relative humidity. So, the variation of thermal conductivity with water content has to be taken into account.

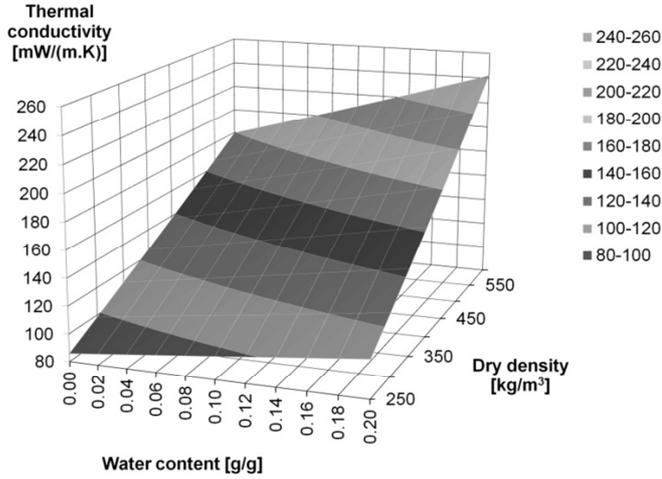

**Figure 11. Thermal conductivity of SHC wall versus dry density and water content.**

Table 3: Thermal conductivity of SHC wall versus dry density and water content

| $\rho_0$ [kg/m³] | 250 | 300 | 350 | 400 | 450 | 500 | 550 | 600 |
|---|---|---|---|---|---|---|---|---|
| n | 0.86 | 0.83 | 0.80 | 0.77 | 0.75 | 0.72 | 0.69 | 0.66 |
| w [g/g] 0.000 | 85.8 | 98.5 | 111.3 | 124.5 | 137.9 | 151.5 | 165.5 | 179.7 |
| 0.040 | 90.2 | 103.8 | 117.7 | 131.9 | 146.4 | 161.2 | 170.9 | 185.7 |
| 0.080 | 94.6 | 109.2 | 124.1 | 139.4 | 155.0 | 170.9 | 176.3 | 191.7 |
| 0.120 | 99.0 | 114.6 | 130.6 | 146.9 | 163.7 | 180.9 | 181.8 | 197.8 |
| 0.160 | 103.4 | 120.1 | 137.1 | 154.6 | 172.5 | 190.9 | 187.3 | 204.0 |
| 0.200 | 107.9 | 125.6 | 143.7 | 162.3 | 181.5 | 201.1 | 192.8 | 210.2 |

## 5   Conclusions

The results presented herein show that the thermal conductivity of hemp concrete depends both on its formulation, its density and its water content. The classical increase of thermal conductivity with density is observed but it is also underlined that the binder type, the hemp shiv type (fibered or defibered) and the hemp to binder ratio deviate the value of thermal conductivity from the variation law. The study of thermal conductivity versus water content and density shows that the water content has a lower effect than density on thermal conductivity (increase lower than 15 to 20% for a wide range of relative humidity). However, this effect should be taken into account in the modelling of hygrothermal behaviour of the building envelope. The thermal conductivity for high density is higher than twice the value for low density with the same formulation.

## Acknowledgments

This article presents results following the French national projet ANR-06-MAPR-0002 Beton chanvre – Programme MAPR 2006 and the Breton regional project "PRIR Ecomatx". These projects were conducted with the financial support of the French National Agency of Research (ANR) and the Brittany Regional Council.

The authors thank the SMEs Easy Chanvre and SI2C for their contribution to the manufacturing of hemp concrete.